\documentclass[a4paper]{article}

\usepackage{INTERSPEECH2022}
\usepackage{algorithm}
\usepackage{algpseudocode}
\usepackage[symbol]{footmisc}
\usepackage{multirow, multicol}
\usepackage{xcolor}

\title{Multi-View Attention Transfer for Efficient Speech Enhancement}

\name{Wooseok Shin\thanks{$^\dagger$These authors contributed equally to this work.}$^\dagger$, Hyun Joon Park$^\dagger$, Jin Sob Kim, Byung Hoon Lee, Sung Won Han\thanks{$^\ast$Corresponding author.}$^\ast$}

\address{School of Industrial and Management Engineering, Korea University, Seoul, Republic of Korea}
\email{\{wsshin95,winddori2002,jinsob,qlscm777,swhan\}@korea.ac.kr}

\begin{document}

\maketitle
\begin{abstract}
Recent deep learning models have achieved high performance in speech enhancement; however, it is still challenging to obtain a fast and low-complexity model without significant performance degradation. Previous knowledge distillation studies on speech enhancement could not solve this problem because their output distillation methods do not fit the speech enhancement task in some aspects. In this study, we propose multi-view attention transfer (MV-AT), a feature-based distillation, to obtain efficient speech enhancement models in the time domain. Based on the multi-view features extraction model, MV-AT transfers multi-view knowledge of the teacher network to the student network without additional parameters. The experimental results show that the proposed method consistently improved the performance of student models of various sizes on the Valentini and deep noise suppression (DNS) datasets. MANNER-S-8.1GF with our proposed method, a lightweight model for efficient deployment, achieved 15.4$\times$ and 4.71$\times$ fewer parameters and floating-point operations (FLOPs), respectively, compared to the baseline model with similar performance.
\end{abstract}
\noindent\textbf{Index Terms}: speech enhancement, multi-view knowledge distillation, feature distillation, time domain, low complexity
\section{Introduction}

Speech enhancement (SE) involves the removal of background noise to improve the perceptual quality of noisy speech. SE is a fundamental task in various speech-processing applications, such as automatic speech recognition, hearing aids, and teleconference systems. Recently, deep learning-based methods have been applied and have significantly improved the performance of SE. These methods can be divided into time domain and time-frequency (TF) domain approaches. TF domain methods use the spectral features obtained by applying the short-time Fourier transform to the signal to estimate clean speech \cite{tan2018convolutional,yin2020phasen,zheng2020interactive,fu2019metricgan,fu2021metricgan+}. In contrast, time domain methods use a raw waveform containing implicitly all of the signal's information to predict clean speech \cite{park2022manner,defossez2020real,wang2021tstnn,pandey2020dual}. 
Although TF and time domain methods have achieved significant improvements in SE, the problem remains that they do not simultaneously satisfy the low computational complexity and model complexity required in various deployment environments while minimizing performance degradation.

To mitigate this issue, knowledge distillation (KD) methods, which transfer useful knowledge from a large teacher network to a compact student network, were first studied in image recognition \cite{hinton2015distilling, romero2014fitnets, zagoruyko2016paying,kim2018paraphrasing}. After verifying that KD can obtain a low-resource student model with satisfactory performance in the image domain, researchers have applied KD methods to obtain efficient student models for speech relevance tasks such as speech recognition \cite{chebotar2016distilling,lu2017knowledge,kurata2020knowledge}, acoustic scene classification \cite{heo2019acoustic,jung2020knowledge}, and acoustic event detection \cite{shi2019compression,shi2019semi}.

In addition, a few studies have exploited the advantages of KD for SE. In the TF domain, \cite{hao2020sub} proposed a sub-band KD framework to improve a single student network using multiple teachers trained for each sub-band. \cite{chen2021light} designed a two-stage training distillation method and a co-worker-based network to improve the performance of SE. In the time domain, to improve performance at both low and high signal-to-noise ratios (SNRs), \cite{hao2020snr} built multiple teachers trained under SNRs and then transferred knowledge to the student network. Further, \cite{nakaoka2021teacher} applied standard KD \cite{hinton2015distilling} to reduce the system latency while preventing performance degradation.

Although the previous KD methods in SE achieved improvements in the model efficiency and performance, they only applied standard KD methods confined to output distillation which is the method to use the soft labels of the teacher as additional supervision of the student. In contrast to standard KD for classification tasks, the disadvantages of standard KD for regression tasks such as SE include the following: \lowercase\expandafter{\romannumeral1}) It cannot take advantage of class distributions (soft labels) in which the student network can easily learn intra-class variations \cite{hinton2015distilling}. \lowercase\expandafter{\romannumeral2}) Because the output of the teacher network is unbounded, it may provide confusing guidance, which may limit performance improvement \cite{kang2021data}. Including the above problems of applying standard KD in SE, the insights that the intermediate features provide richer information than outputs \cite{romero2014fitnets, kim2018paraphrasing, tung2019similarity} suggest the necessity of feature-based KD in SE. In addition, recent SE models \cite{park2022manner,wang2021tstnn, pandey2020dual} based on a dual-path method that extracts local and global contextual information from long-range speech waveforms motivated us to design feature-wise knowledge transfer.

In this study, we propose multi-view attention transfer (MV-AT) to obtain efficient speech enhancement models. To the best of our knowledge, this is the first feature-based KD method for time domain SE. We adopted attention transfer (AT) \cite{zagoruyko2016paying} applying the same transformation to the features of the teacher and student for feature-based distillation as the base feature transformation method. As studied in MANNER \cite{park2022manner}, which describes the importance of separated signal feature representations (i.e., channel, local, and global), MV-AT based on the MANNER backbone transfers the feature-wise knowledge of the teacher by utilizing each feature highlighted in the multi-view. By applying MV-AT, the student network can easily learn the teacher’s signal representation and mimic the matched representation from each perspective. Furthermore, MV-AT can not only compensate for the limitation of standard KD in SE as feature-based distillation but also make an efficient SE model without additional parameters. The results of experiments conducted on the Valentini and DNS datasets indicate that the proposed method achieves significant efficiency. While exhibiting comparable performance to the baseline model, the model generated by the proposed method required 15.4$\times$ and 4.71$\times$ fewer parameters and flops, respectively.
%
\begin{figure*}[t]
  \centering
  \includegraphics[scale=0.65]{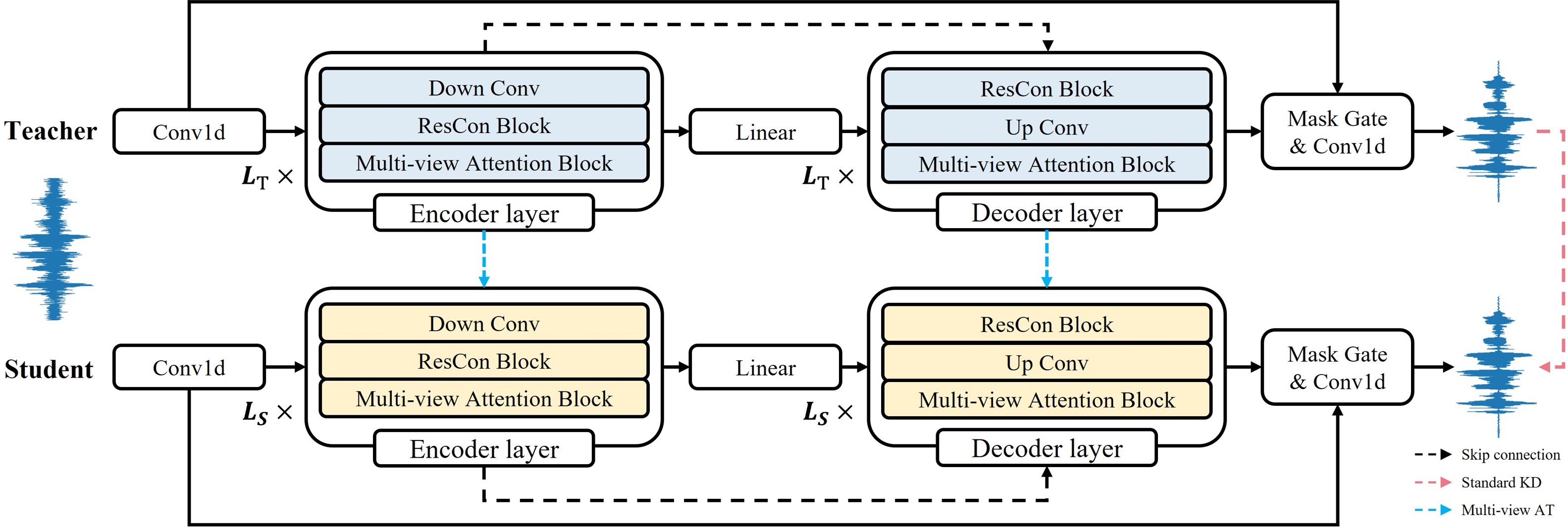} 
  \caption{Proposed time domain distillation framework based on the MANNER architecture.} 
  \label{fig:architecture}
\end{figure*}
\section{Method}
This section reviews the MANNER \cite{park2022manner} architecture used as the base architecture in this study and presents our MV-AT method based on feature distillation.
%
\subsection{Base Architecture}
We adopted a state-of-the-art time domain SE model, MANNER, as the base architecture of the teacher and student network for knowledge distillation. As shown in Figure~\ref{fig:architecture}, MANNER consists of a multi-layer ($L_T$, $L_S$) encoder and decoder with skip connections. The encoder and decoder contain Down and Up Conv layers, a residual conformer (ResCon) block, and a multi-view attention (MA) block. Given a noisy input $x$, a 1D convolution expands the channel size of $x$. The latent representations are extracted by the encoder applied to $x$, followed by a linear transformation, and are passed to the decoder. While forwarding $x$, the MA block in each encoder and decoder layer emphasizes each channel, global, and local signal representation to consider all the signal information. Finally, the enhanced output $\hat{y}$ is obtained by applying a mask gate followed by convolution. To obtain an efficient student network, we set the student network to be composed of smaller depth, channel dimension, and the number of MA blocks than the teacher network, reducing computational and model complexity.
%
\newcommand\normx[1]{\left\Vert#1\right\Vert}   
\subsection{Attention Transfer} \label{section:AT}
To utilize AT in time domain SE, we define an activation map $A^T \in \mathbb{R}^{C\times t}$ of a pre-trained teacher network (T) and the corresponding activation map $A^S \in \mathbb{R}^{C\times t}$ of a student network (S). Here, $C$ and $t$ denote the channel dimension and signal length. Subsequently, a mapping function $\mathcal{F}$ receives the activation map (i.e., $A^T$ or $A^S$) as an input, aggregates it in the channel dimension, and outputs a temporal attention map (TAM) as follows:
\begin{equation}
  \mathcal{F}(A) = \Sigma_{i=1}^{C} |A_i|^p \;\; \in \mathbb{R}^{1\times t}
  \label{eq1}
\end{equation}
where $A_i$ denotes the activation map for the $i$\textsuperscript{th} channel and $p$, which adjusts the intensity of the TAM, is set to two, as in \cite{zagoruyko2016paying}. AT assumes that the absolute value of the hidden activation can be used as an indicator of importance. Based on this, as shown in Eq.~\ref{eq1}, we extract the TAM for transferring knowledge by aggregating channel representations. Then, $\mathcal{F}(A)$ is replaced by the $l_2$-normalized form $\frac{\mathcal{F}(A)}{\normx{\mathcal{F}(A)}_2}$, as done in \cite{zagoruyko2016paying}. We define the AT loss for a pair of teacher-student attention maps using $l_1$ loss (p=1) as follows below. (It is also possible to use the $l_2$ (p=2) loss, but we used the $l_1$ loss in all experiments because $l_1$ is slightly better than $l_2$. See Section \ref{section_ablation}.)
\begin{equation}
  \mathcal{L}_{AT} = \normx{ \frac{\mathcal{F}(A^T)}{\normx{\mathcal{F}(A^T)}_2} - \frac{\mathcal{F}(A^S)}{\normx{\mathcal{F}(A^S)}_2} }_p
  \label{eq2}
\end{equation}
\begin{figure}[t]
  \centering
  \includegraphics[width=\linewidth]{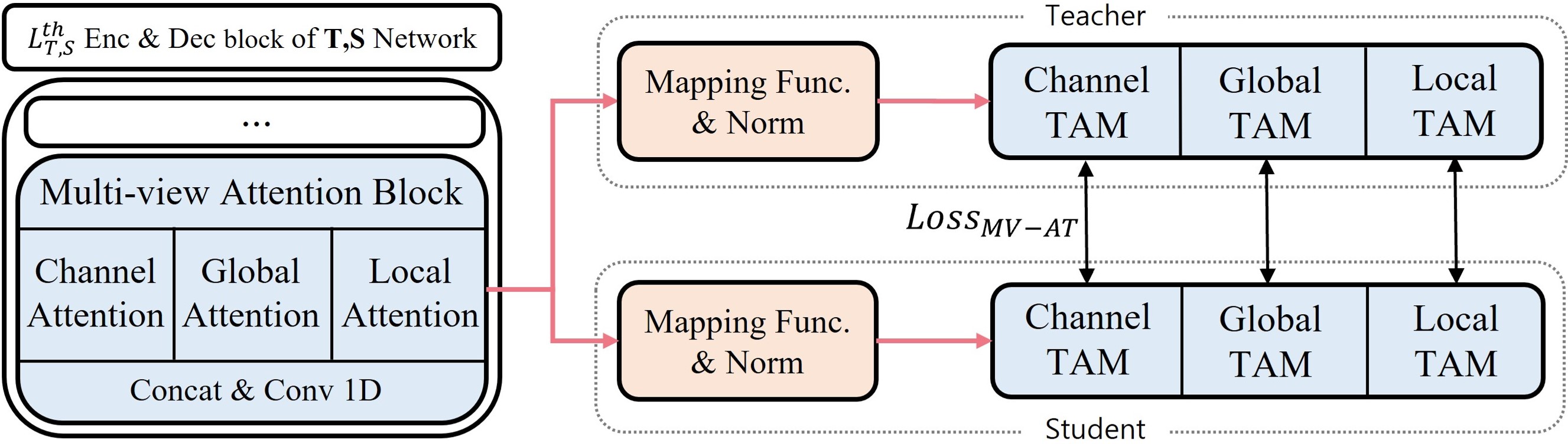}
  \caption{Flow of multi-view attention transfer. $L_{T,S}^{th}$ indicates the layer of teacher and student.}
  \label{fig:MVAT}
\end{figure}
%
\subsection{Multi-View Attention Transfer}
As mentioned above, we adopt the MANNER architecture based on the MA block to transfer the diverse knowledge emphasized in the multi-view. Figure~\ref{fig:MVAT} presents the process of obtaining temporal attention maps (TAMs) from the MA blocks of the encoder and decoder layers. After each attention in the MA block, the mapping function $\mathcal{F}$ is applied to the three representations, followed by normalization. The MV-AT loss is then defined for a teacher-student pair with three TAMs each.
\begin{equation}
  \mathcal{L}_{MV-AT} = \Sigma_{view \in \{c,g,l\}} \;\, \mathcal{L}_{AT_{view}}
  \label{eq3}
\end{equation}
where $c$, $g$, and $l$ denote channel, global, and local, respectively. This allows the student network to mimic the manner in which the teacher network emphasizes each view of the signal representations. For the total distillation loss, we combine MV-AT with the standard KD loss distilling the network output. The loss is defined as follows:
\begin{equation}
  \mathcal{L}_{distill} = \Sigma_{j \in \mathcal{I}} \;\, \mathcal{L}^{j}_{MV-AT} + \mathcal{L}_{KD}(\hat{y}_T, \hat{y}_S)
  \label{eq4}
\end{equation}
where $\mathcal{I}$ indicates the index of all teacher-student activation layer pairs for which knowledge is transferred. 
Regarding $\mathcal{L}_{KD}$, we adopt a weighted loss considering the estimated clean and noisy speech, where each loss consists of $l_1$ loss and multi-resolution STFT loss, as done by \cite{park2022manner}.
%
\begin{table*}[th]
\caption{Objective evaluation results of the proposed method on the Valentini test dataset.}
\label{table:main_result}
\centering
\resizebox{0.85\linewidth}{!}{%
\begin{tabular}{c|ccccccc}
\toprule
Model & Params (M) & FLOPs (G) & PESQ & STOI(\%) & CSIG & CBAK & COVL \\
\midrule
{MANNER-B-Teacher \hspace{0pt plus 1filll}} & 24.07 & 61.92 & 3.21 & 95 & 4.53 & 3.65 & 3.91 \\
\midrule
\midrule
{MANNER-S-1.5GF \, (Baseline) \hspace{0pt plus 1filll}} & 0.26 & 1.46 & 2.88 & 94 & 4.32 & 3.47 & 3.62\\
{MANNER-S-1.5GF + MV-AT \hspace{0pt plus 1filll}} & 0.26 & 1.46 & 2.94 & 94 & 4.34 & 3.49 & 3.67\\
{MANNER-S-1.5GF + MV-AT + KD \hspace{0pt plus 1filll}} & 0.26 & 1.46 & 2.95 & 94 & 4.33 & 3.50 & 3.66\\
{DEMUCS-S \cite{defossez2020real} \hspace{0pt plus 1filll}} & 18.87 & 34.56 & 2.93 & 95 & 4.22 & 3.25 & 3.52\\
{TSTNN \cite{wang2021tstnn} \hspace{0pt plus 1filll}} & 0.92 & 166.54 & 2.96 & 95 & 4.33 & 3.53 & 3.67\\
\midrule
{MANNER-S-5.3GF \, (Baseline) \hspace{0pt plus 1filll}} & 0.90 & 5.32 & 2.98 & 95 & 4.38 & 3.54 & 3.71\\
{MANNER-S-5.3GF + MV-AT \hspace{0pt plus 1filll}} & 0.90 & 5.32 & 3.04 & 95 & 4.42 & 3.56 & 3.76\\
{MANNER-S-5.3GF + MV-AT + KD \hspace{0pt plus 1filll}} & 0.90 & 5.32 & 3.06 & 95 & 4.42 & 3.58 & 3.77\\
{DEMUCS-L \cite{defossez2020real} \hspace{0pt plus 1filll}} & 33.53 & 388.48 & 3.07 & 95 & 4.31 & 3.40 & 3.63 \\
\midrule
{MANNER-S-8.1GF \, (Baseline) \hspace{0pt plus 1filll}} & 1.38 & 8.14 & 3.01 & 95 & 4.40 & 3.55 & 3.74\\
{MANNER-S-8.1GF + MV-AT \hspace{0pt plus 1filll}} & 1.38 & 8.14 & 3.07 & 95 & 4.43 & 3.58 & 3.78\\
{MANNER-S-8.1GF + MV-AT + KD \hspace{0pt plus 1filll}} & 1.38 & 8.14 & 3.12 & 95 & 4.45 & 3.61 & 3.82\\
{MANNER-S-38.GF \hspace{0pt plus 1filll}} & 21.25 & 38.34 & 3.12 & 95 & 4.45 & 3.60 & 3.81 \\
\bottomrule
\end{tabular}
}
\end{table*}
\subsection{Dual-depth Matching}
We assume that different depths contain different knowledge and feature distillation of the same location allows the student network to easily mimic the teacher network. If $L_S$ and $L_T$ are the same, the teacher-student features of the same depth position are matched one-to-one when transferring the feature knowledge. By contrast, for efficiency, when $L_S$ is smaller than $L_T$, we utilize a dual-depth knowledge transfer strategy. The basis for this is that a large teacher network can extract sufficiently detailed representations, whereas a student network only learns representative representations in a compact space. 
For example, if $L_T^{i_{\in \{1-4\}}}$ and $L_S^{j_{\in \{1-3\}}}$ are 4 and 3 layer encoder-decoder, the student's third layer $L_S^3$ contains partial representations of the teacher's third and fourth layers ($L_T^3$ and $L_T^4$). Thus, we distill both the third- and fourth-layer features of the teacher adjacent to the third-layer feature of the student, called dual-depth knowledge transfer. We adopt interpolation to match the signal lengths of teacher-student attention maps under different layer pairs when we utilize the dual-depth knowledge transfer.
%
\begin{table*}[th]
\caption{Objective evaluation results of the proposed method on the DNS no reverb test dataset.}
\label{table:dns_result}
\centering
\resizebox{0.85\linewidth}{!}{%
\begin{tabular}{c|ccccccc}
\toprule
Model & Params (M) & FLOPs (G) & PESQ & STOI(\%) & CSIG & CBAK & COVL\\
\midrule
{MANNER-S-Teacher \hspace{0pt plus 1filll}} & 21.25 & 38.34 & 3.07 & 97.1 & 4.52 & 3.81 & 3.84\\
\midrule
\midrule
{MANNER-S-1.5GF \, (Baseline) \hspace{0pt plus 1filll}} & 0.26 & 1.46 & 2.47 & 94.8 & 4.07 & 3.35 & 3.27\\
{MANNER-S-1.5GF + MV-AT \hspace{0pt plus 1filll}} & 0.26 & 1.46 & 2.47 & 94.9 & 4.07 & 3.34 & 3.27\\
{MANNER-S-1.5GF + MV-AT + KD \hspace{0pt plus 1filll}} & 0.26 & 1.46 & 2.47 & 94.9 & 4.04 & 3.34 & 3.25\\
\midrule
{MANNER-S-5.3GF \, (Baseline) \hspace{0pt plus 1filll}} & 0.90 & 5.32 & 2.70 & 95.9 & 4.25 & 3.53 & 3.48\\
{MANNER-S-5.3GF + MV-AT \hspace{0pt plus 1filll}} & 0.90 & 5.32 & 2.75 & 96.1 & 4.30 & 3.58 & 3.55\\
{MANNER-S-5.3GF + MV-AT + KD \hspace{0pt plus 1filll}} & 0.90 & 5.32 & 2.72 & 96.0 & 4.25 & 3.53 & 3.49\\

\bottomrule
\end{tabular}
}
\end{table*}
\section{Experiments}
\subsection{Experimental Setup}
\textbf{Implementation details} In experiments conducted on the Valentini and DNS datasets, we used MANNER-Base (B) and MANNER-Small (S), respectively, as powerful teacher networks. The depth and channel dimensions of the teacher networks were set to 4 and 60 to transfer the knowledge of enriched representations. We employed MANNER-S, which includes an MA block only in the $L\textsuperscript{th}$ layer, as a compact student network for efficiency. In addition, small depths and channel dimensions were used to reduce the complexity without changing the structure of the base architecture. To compare student models at various complexities, we suffixed the flop regime to the models (e.g., MANNER-S-1.5GF). Specifically, the MANNER-S-1.5GF, 5.3GF, and 8.1GF models had depths of 3 and channel dimensions of 12, 24, and 30, respectively. Here, GF indicates $10^9$ flops. For training settings, we set a batch size of 4 and 350 epochs to train the student networks. Except for the smaller depth, channel dimension, and the number of MA blocks than those of the teacher networks for achieving high efficiency of the student networks, we adopted the other parameters and learning strategies the same as those of MANNER.

\noindent\textbf{Datasets}
To compare the proposed method with previous methods, we used the Valentini dataset \cite{valentini2017noisy}, for which the total time is approximately 10 h. The dataset consists of 30 speakers; we used 28 speakers in the train and validation sets, and the remainder in the test set. The train set comprised 11,572 utterances with four SNRs (15, 10, 5, and 0 dB) of noise. The test set comprised 824 utterances with four SNRs (17.5, 12.5, 7.5, and 2.5 dB) of unseen noise.

To evaluate the proposed method on a larger dataset, we adopted the DNS dataset \cite{reddy2020interspeech}. For training, we generated 50 h of utterances from the clean speech set and mixed them with noise data at various SNRs. Among the given test sets, we selected the test set that did not contain reverberations for the evaluation. The SNRs of the noise in the test set were distributed between 0 dB and 20 dB. All datasets used were downsampled from 48 to 16 kHz.

\noindent\textbf{Evaluation metrics}
We evaluated the models in terms of objective measures for the enhanced speech. For objective measures that are used to evaluate speech quality and intelligibility, we adopted the perceptual evaluation of speech quality (PESQ) \cite{recommendation2001perceptual} and short-time objective intelligibility (STOI) \cite{taal2011algorithm}. The PESQ score ranges from -0.5 to 4.5, and that of STOI ranges from 0 to 100. As composite measures, we adopted CSIG, CBAK, and COVL \cite{hu2007evaluation}, which are used to evaluate signal distortion, noise intrusiveness, and overall signal quality, respectively. The scores on the three measures range from 1 to 5. Furthermore, we recorded the floating-point operations (FLOPs) and the number of model parameters to evaluate model efficiency.
\subsection{Results}
To verify the efficacy of the proposed method, we conducted experiments using different base network configurations on the Valentini and DNS datasets. We compared the performance of three training setups: Base model, Base model+MV-AT, and Base model+MV-AT+KD, where ``MV-AT'' is the proposed method and ``KD'' is the standard KD \cite{hinton2015distilling} that distills the output. The top row of each table represents the teacher network. As demonstrated in Table~\ref{table:main_result}, the proposed method achieved significant performance improvements without increasing the parameters and flops in the evaluation metrics compared to the baseline. The PESQ scores of the model to which MV-AT was applied were 0.06 higher than the baseline scores. Moreover, the combination of MV-AT and KD improved the overall performance slightly more than when MV-AT was applied as a standalone. This result suggests that the intermediate features and output of the teacher network have different knowledge and synergize knowledge transfer.

In terms of efficiency, the proposed method was compared with the state-of-the-art methods DEMUCS~\cite{defossez2020real} and TSTNN~\cite{wang2021tstnn} achieving comparable performance. MANNER-S-1.5GF (with MV-AT and KD) outperformed DEMUCS-S while requiring 72.58$\times$ and 23.67$\times$ fewer parameters and flops, respectively. Compared to TSTNN, MANNER-S-1.5GF (with MV-AT and KD) showed similar performance; however, MANNER-S-1.5GF required 3.54$\times$ and 114.07$\times$ fewer parameters and flops, respectively, than TSTNN. Similarly, MANNER-S-5.3GF (with MV-AT and KD) has 37.2$\times$ and 73.02$\times$ fewer parameters and flops than DEMUCS-L, respectively, while giving comparable performance. For efficiency comparison within the same structure (MANNER), we compared MANNER-S-8.1GF (with MV-AT and KD) to MANNER-S-38GF. While giving comparable performance, MANNER-S-8.1GF (with MV-AT and KD) achieved 15.4$\times$ and 4.71$\times$ fewer parameters and flops, respectively, than MANNER-S-38GF.

We also conducted an experiment to validate the effectiveness of the proposed method on a larger DNS dataset. Table~\ref{table:dns_result} presents the objective evaluation results of the no-reverb test dataset. Interestingly, MANNER-S-1.5GF, the most compact network, did not benefit from the proposed method, whereas MANNER-S-5.3GF showed an overall performance improvement. This result suggests that it is difficult for a student network with only 0.26M parameters to learn the vast knowledge of the teacher network obtained from a large dataset. Furthermore, in contrast to the results obtained on the Valentini dataset, we observed that MANNER-S-5.3GF with MV-AT and KD showed limited performance improvement compared to MANNER-S-5.3GF with MV-AT standalone. As a result, our proposed MV-AT achieved a consistent performance improvement in all cases, suggesting that it is a suitable knowledge distillation method for SE tasks.
\begin{table}[th]
\caption{Ablation study for the MANNER-S-1.5GF model with L=3, C=12 using the Valentini dataset. ${\dagger}$ indicates that dual-depth knowledge transfer is employed.}
\label{table:ablation}
\centering
\resizebox{\linewidth}{!}{%
\begin{tabular}{c|ccccc}
\toprule
Model & PESQ & CSIG & CBAK & COVL \\
\midrule
\midrule
{Baseline (w/o distill.) \hspace{0pt plus 1filll}} & 2.88 & 4.32 & 3.47 & 3.62\\
{Single-view AT$^{\dagger}$ \hspace{0pt plus 1filll}} & 2.90 & 4.32 & 3.48 & 3.63\\
{KD \hspace{0pt plus 1filll}} & 2.91 & 4.33 & 3.49 & 3.64\\
{MV-AT (w/o dual-depth) \hspace{0pt plus 1filll}} & 2.92 & 4.33 & 3.49 & 3.65\\
{MV-AT$^{\dagger}$ ($l_1$-norm) \hspace{0pt plus 1filll}} & 2.94 & \textbf{4.34} & 3.49 & \textbf{3.67}\\
{MV-AT$^{\dagger}$ ($l_2$-norm) \hspace{0pt plus 1filll}} & 2.93 & 4.33 & 3.49 & 3.65\\
{MV-AT$^{\dagger}$ + KD \hspace{0pt plus 1filll}} & \textbf{2.95} & 4.33 & \textbf{3.50} & 3.66\\
\bottomrule
\end{tabular}
}
\end{table}
\subsection{Ablation Study} \label{section_ablation}
To investigate the effects of different components on the performance of the proposed method, we performed an ablation study over MV-AT and standard KD. The results in Table~\ref{table:ablation} show that both the standard KD and feature-based distillation improved the performance of the student network. To validate the influence of the proposed MV-AT, we evaluated the effect of replacing MV-AT with a single-view (standard) AT that distills the output features of each encoder-decoder block. Although the network adopting single-view AT outperformed the baseline, it exhibited limited performance improvement compared to MV-AT. This suggests that distilling the multi-view representations works complementarily, making it easier for compact student networks to converge. We also observed a slight improvement in performance when using the standard KD method alone. However, under single-component conditions, the proposed MV-AT significantly outperformed other methods. Moreover, the MV-AT with dual-depth knowledge transfer performed better than the single-depth MV-AT. As mentioned in Section \ref{section:AT}, we used the $l_1$ (p=1) loss to compute the difference between the teacher and student attention maps. We also explored which p-values tended to work better for p=1 and p=2. Both cases achieved similar performance, but because the $l_1$ loss was slightly better than the $l_2$ loss, we adopted the $l_1$ distance loss.
\section{Conclusions}
In this study, we proposed multi-view attention transfer (MV-AT), which distills features highlighted from three perspectives in speech enhancement tasks that require low complexity for deployment in various environments. Experimental results show that the proposed method achieves significant efficiencies by reducing the model parameters and computational complexity by 15.4$\times$ and 4.71$\times$, respectively, compared to the baseline. An ablation study suggests that transferring diverse knowledge, including multi-view features and output, significantly improves performance. Finally, further experiments on a large DNS dataset suggest that MV-AT is a suitable knowledge distillation method for speech enhancement tasks.
\section{Acknowledgements}
This research was supported by Brain Korea 21 FOUR. This research was also supported by Korea University Grant (K2202151).

\bibliographystyle{IEEEtran}

\bibliography{main}

\end{document}